# Enthalpic and entropic phase transitions in high energy density nuclear matter

## Igor Iosilevskiy

*JIHT RAS; MIPT, Moscow, Russia*

iosilevskiy@gmail.com

Features of Gas-Liquid (GL) and Quark-Hadron (QH) phase transitions (PT) in dense nuclear matter are under discussion in comparison with their terrestrial counterparts, e.g. so-called "plasma" PT in shock-compressed hydrogen, nitrogen etc. Both, GLPT and QHPT, when being represented in widely accepted $T - \mu$ plane, are often considered as similar, i.e. amenable to one-to-one mapping by simple scaling. It is argued that this impression is illusive and that GLPT and QHPT belong to different classes: GLPT is typical enthalpic (VdW-like) PT while QHPT ("deconfinement-driven") is typical entropic PT (like hypothetical ionization- and dissociation-driven phase transitions in dense hydrogen, nitrogen etc. in megabar pressure range). Fundamental differences of enthalpic and entropic phase transitions are discussed and illustrated.

Phase transition (PT) is universal phenomena in many terrestrial and astrophysical applications. There are very many variants of hypothetical PTs in ultra-high energy and density matter in interiors of neutron stars (so-called hybrid or quark-hadron stars) [1], in core-collapse supernovae explosions and in products of relativistic ions collisions in modern super-colliders (LHC, RHIC, FAIR, NICA etc.). Two hypothetical 1$^{st}$-order phase transitions are the most widely discussed in study of high energy density matter ($\rho \sim 10^{14}$ g/cc): (***i***) – gas-liquid-like phase transition (GLPT) in ultra-dense nuclear matter: i.e. in equilibrium (Coulombless) ensemble of protons, neutrons and their bound clusters $\{p, n, N(A,Z)\}$ at $T \leq 20$ MeV, and (***ii***) – quark-hadron (deconfinement) phase transition (QHPT) at $T \leq 200$ MeV. (see e.g. [2]). Both, GLPT and QHPT, when being represented in widely accepted $T – \mu$ plane ($\mu$ – baryonic chemical potential) are often considered as similar, i.e. amenable to one-to-one correspondence with possible transformation into each other by simple scaling (see e.g. Figs.1 and 12 in [3]). The main statement of present paper is that this impression is illusive and that GLPT and QHPT belong to different classes: GLPT is typical *enthalpic* (VdW-like) PT, while "deconfinement-driven" QHPT is typical *entropic* PT like hypothetical ionization- and dissociation-driven phase transitions in shock-compressed dense hydrogen, nitrogen etc. in megabar pressure range (see e.g. [4]). Fundamental differences of these two types of PT are discussed and illustrated below.

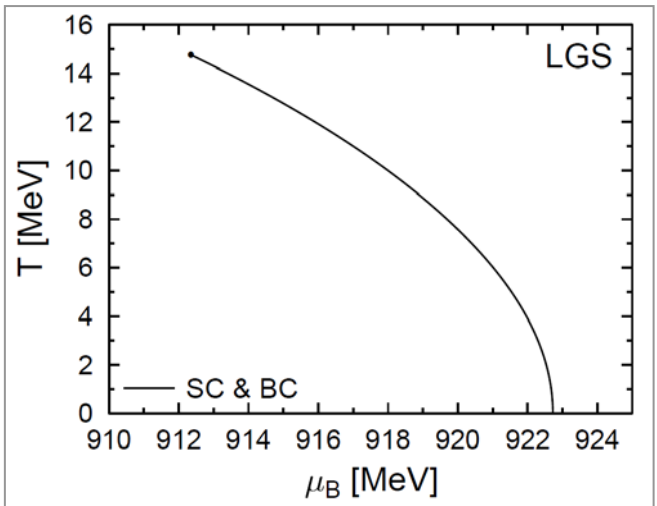

**Figure 1**. Gas-liquid phase transition in symmetric nuclear matter $\{p,n,N(A,Z)\}$ in temperature – barion chemical potential plane. (FSUGold RMF model. Figure from [3]).

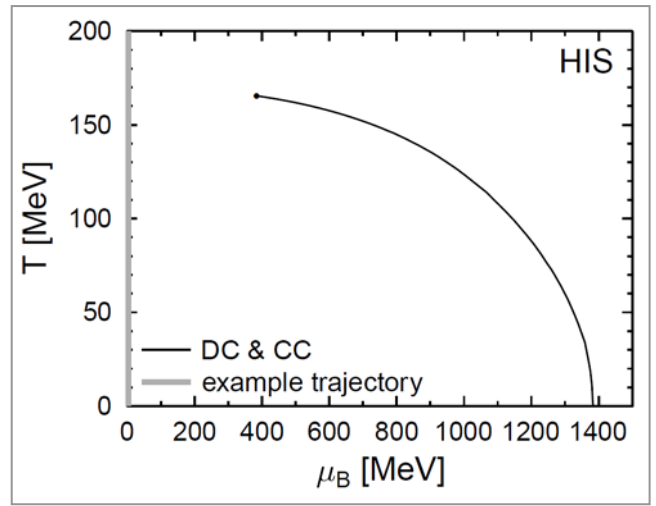

**Figure 2**. Quark-hadron (deconfinement) phase transition (QHPT) in temperature – barion chemical potential plane. (SU(3) model. Figure from [3])

## Comparison of GLPT and QHPT in density-temperature plane

GLPT and QHPT look as similar in $T$ - $\mu$ plane (Figs. 1, 2). It should be noted that unfortunately this type of representation is not revealing for PT analysis. Fundamental difference between GLPT and QHPT could be more evidently demonstrated in other variants of phase diagram widely used in electromagnetic plasmas community (see e.g. [4]). First one is density–temperature ($T – \rho$) diagram. Two these phase transitions (GLPT and QHPT) are sometimes considered in $T – \rho$ plane as quantitatively, not qualitatively different in their schematic comparison (see e.g. Fig.2 in [6] and slide 2 in [5]). Numerical calculations of phase boundaries for GLPT and QHPT (see Figures 3 and 14 in [3]) demonstrate significant difference in structure of these two boundaries (Figure 3 below [HV]).

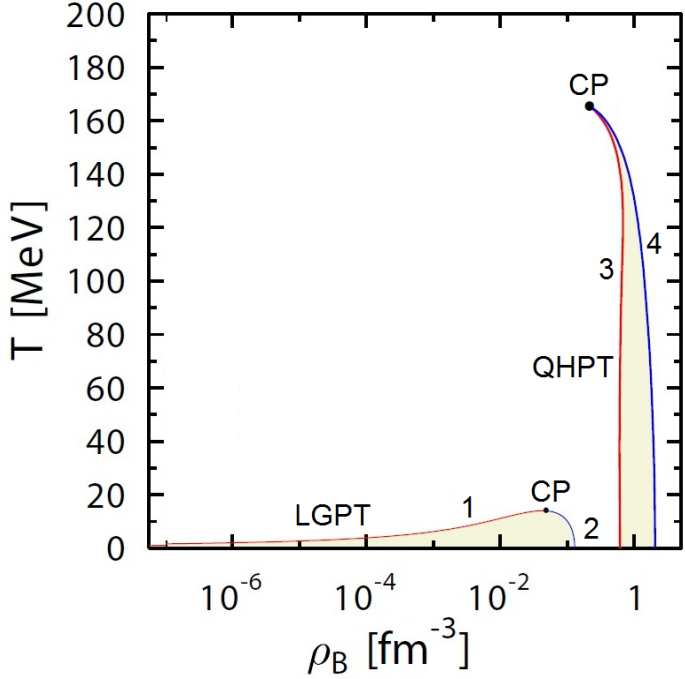 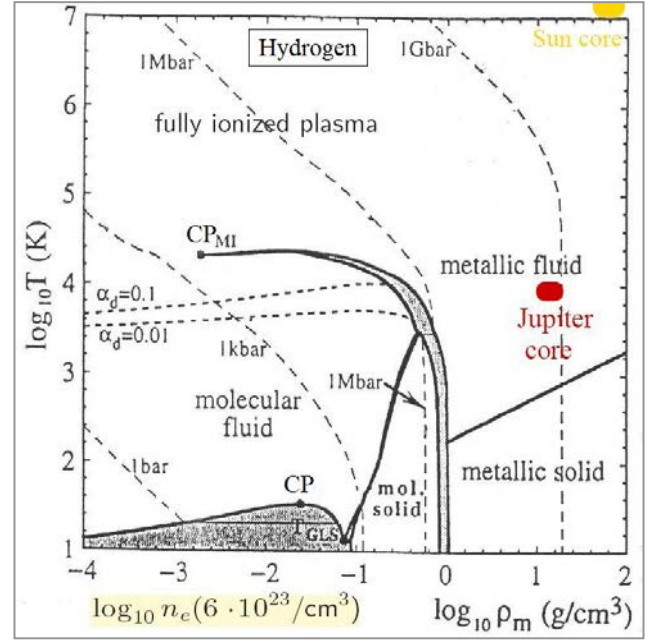

**Figure 3**. Gas–liquid and quark–hadron phase transitions (GLPT vs. QHPT) in high energy and density range [3,7]. *Phase boundaries*: *1* – saturation, *2* – boiling, *3* – deconfinement, *4* – hadronization, *CP* – critical points (1 fm$^{-3}$ = 10$^{39}$ cm$^{-3}$; 1 Mev $\Leftrightarrow$ 1.16$*$10$^{10}$ K).

**Figure 4**. Gas-liquid and plasma phase transitions (GLPT and PPT) in hydrogen (Figure from [8]). *Phase boundaries* (left to right): GLPT – saturation, boiling and sublimation; PPT – pressure ionization; *CP* – critical points.

It should be stressed [9] that remarkably similar structure of boundaries for two phase transitions are well known in electromagnetic plasmas. For example it is gas-liquid (left) and ionization-driven (right) phase transitions in dense hydrogen (Fig.4) (see Figure in [8]).

## Enthalpic and entropic phase transitions

It is almost evident that two gas-liquid-like PTs, from one side, and two "delocalozation-driven" PTs (QHPT and PPT), from other side, are similar to each other. This similarity in forms of phase boundaries manifests identity in key physical processes, which rule by phase transformations in both systems in spite of great difference in their densities and temperatures. When one compress isothermally "vapor" phase (subscript *V*) in case of GLPT, he reaches saturation conditions. At this moment the system jumps into "liquid" phase (subscript *L*) with *decreasing* of enthalpy and increasing of nega-entropy in accordance with equality rule for Gibbs free energy in 1$^{st}$-order PT (1)(2).

$$G_V = H_V - TS_V = H_L - TS_L = G_L \tag{1}$$

$$\Delta G = 0 \quad \Leftrightarrow \quad \Delta H = H_V - H_L = T(S_V - S_L) \geq 0 \tag{2}$$

Opposite order of enthalpy and entropy change should be stressed for both "delocalozation-driven" phase transitions (QHPT and PPT) in Figs. 1 and 2. The both systems, molecular hydrogen (M) and hadronic mixture (H), are ensembles of bound clusters, composed from "elementary" particles: protons and electrons in the case of hydrogen and u- and d-quarks in the case of QHPT. The both systems reaches "pressure-deconfinement" and "pressure-ionization" conditions under iso–$T$ compression and then jump into deconfinement (Q) and plasma (P) phases correspondingly with *decreasing* enthalpy and *increasing* nega-entropy (3), which is just opposite to (2):

$$\Delta G_{\text{PPT}} = 0 \quad \Leftrightarrow \quad \Delta H = H_P - H_M = T(S_P - S_M) \geq 0 \qquad (3)$$

$$\Delta G_{\text{QHPT}} = 0 \quad \Leftrightarrow \quad \Delta H = H_Q - H_H = T(S_Q - S_H) \geq 0 \qquad (3^*)$$

Here indexes "M" vs. "P" and "H" vs. "Q" denote "bound" and "non-bound" phases: molecular vs. plasma, and hadron vs. quark phases correspondingly. It is well-known that quark-gluon plasma (QGP) has "much greater number for degrees of freedom" than hadronic phase (see e.g. [2]). It just means much higher entropy of QGP vs. hadronic phase in thermodynamic terms. This opposite order of enthalpy and entropy changes in two discussed above phase transformation (GLPT and QHPT) is main reason for phase transition classification and terminology accepted and proposed in present paper: namely *enthalpic* (GLPT) vs. *entropic* (QHPT and PPT) phase transitions.

It is evident that besides well-known ionization-driven (plasma) PT, there are many other candidates for being members of entropy transitions class, namely those PTs, where *delocalization* of bound complexes is ruling mechanism for phase transformation. It is e.g. well-known *dissociation-driven* PT in dense hydrogen, nitrogen and other molecular gases (e.g. [10–12] etc.); more exotic *polimerization-* and *depolimerization-driven* PTs in dense nitrogen and possibly other molecular gases (e.g. [13–16] etc.). In all these cases basic feature of entropic (3) and enthalpic (2) PTs leads immediately to opposite sign of $P(T)$–dependence at phase coexistence curve in accordance with Clausius – Clapeiron relation. Hence the slope of $P(T)_{\text{binodal}}$ is the key feature for distinguish both types of PTs, enthalpic (4) and entropic (5).

$$\Delta H = T\Delta S > 0 \quad \Rightarrow \quad (dP/dT)_{\text{binodal}} > 0 \qquad \text{(enthalpic PT)} \qquad (4)$$

$$\Delta H = T\Delta S < 0 \quad \Rightarrow \quad (dP/dT)_{\text{binodal}} < 0 \qquad \text{(entropic PT)} \qquad (5)$$

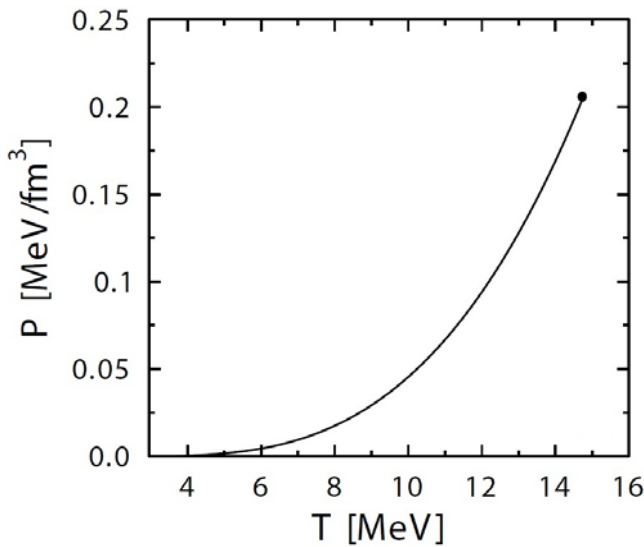 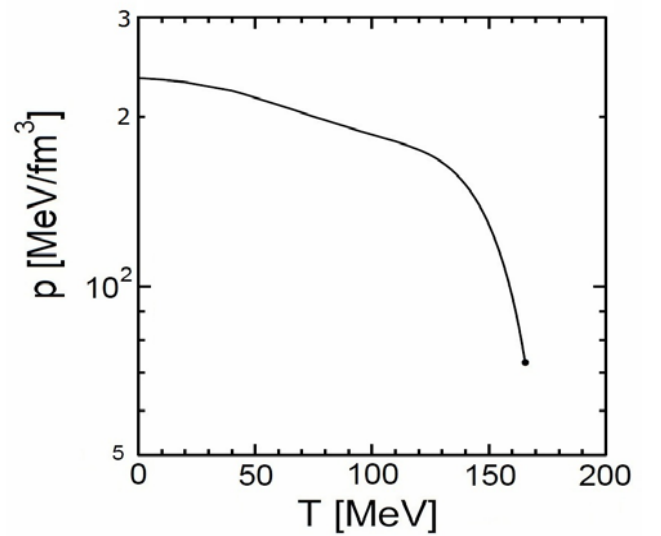

**Figure 5**. *P–T* phase diagram of gas-liquid phase transition in dense symmetric nuclear matter (FSUGold RMF model. Figure from [3]).

**Figure 6**. *P–T* phase diagram of quark-hadron phase transition in symmetric quark-hadron system (Figure from [3]).

## Comparison of GLPT and QHPT in pressure-temperature plane

Exponentially increasing (VdW-like) form of *P–T* phase diagram for ordinary GLPT in hydrogen and other substances is well-known. Similar *P-T* dependence of GLPT in nuclear matter was calculated many times, e.g. [19] [18] [3] etc (see e.g. Figure 5). In contrast to that *P–T* phase diagram of QHPT (see Figure 6) is not widely known ([17] [6] [5]). It was calculated recently in [3]. Both phase transitions, GLPT and QHPT, have opposite $P(T)$ dependence in agreement with (4) and (5).

## Comparison of GLPT and QHPT in pressure-density plane

The most striking difference between enthalpic vs. entropic types of phase transitions could be demonstrated in comparison of their *P–V* phase diagrams. This type of phase diagram is very important for analysis of main dynamic processes in dense plasma: e.g. shock or isentropic compression and adiabatic expansion. *P–V* phase diagram for VdW-like GLPT in ordinary substances is well known. GLPT in symmetric Coulombless nuclear matter has the same structure (see e.g. [19] [18] etc.). In contrast to that *P–V* phase diagram for QHPT was not explored yet. But it will be done soon on the base of QHPT calculations in [3].

Good example of *P–V* phase diagram for entropic ("plasma") phase transition (PPT) in Xenon is exposed at Figure 7 accepted from [20] (see also Fig. III.6.11a in [4]).

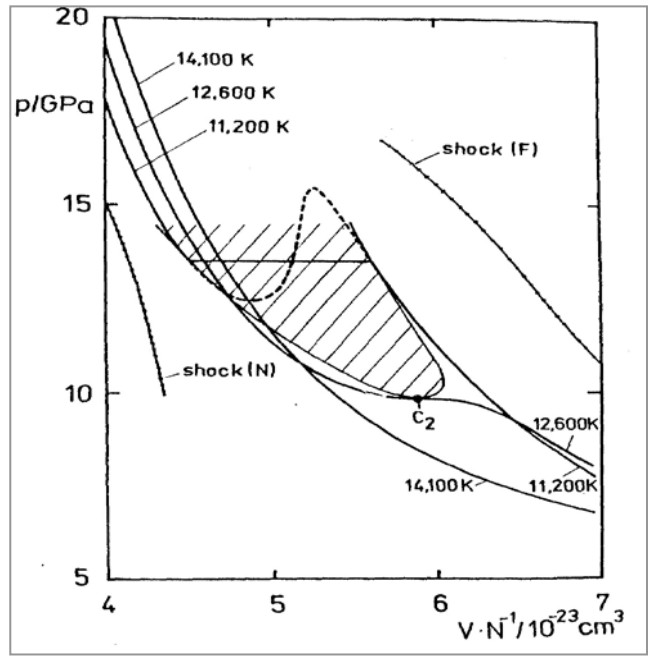

**Figure 7**. *P–V* phase diagram of hypothetical ionization-driven ("plasma") phase transition in xenon. Solid lines – calculated isotherms for $T < T_c$, $T > T_c$ and $T = T_c \approx 12\,600$ K. Initial VdW-loop and equilibrium part of tw0-phase isotherm are shown for $T = 11\,200$ K. Shaded area – two-phase region. $C_2$ – critical point. Dot-solid lines – estimated parameters of shock adiabats; *N* – Nellus *et al.*, *F* – Fortov *et al*. (Figure from [20]).

Several important features of anomalous thermodynamic behavior in two-phase region of this PPT and its close vicinity were demonstrated at the Figure 7 from [20]:

(1) – isotherms *T* =10 – 15 kK (below and above critical isotherm) cross each other;

(2) – low-*T* isotherm lay above high-*T* ones. Features (*1-2*) means *negative sign* of thermal pressure coefficient $(\partial P/\partial T)_V < 0$ in discussed region;

(3) – one meets anomalous behavior of isotherms within two-phase region of this PPT at sufficiently low temperature, namely:
  - appearance of return-point in spinodal region on low-density part of isotherm;

- appearance of *third* metastable section with positive commressibility (i.e. $(\partial P/\partial V)_T < 0$) between two *unstable* parts of isotherm within spinodal region.

Features (2) and (3) are in contrast to standard behavior of gas-liquid PT, where one unstable part of isotherm divides two metastable parts in ordinary VdW-loop.

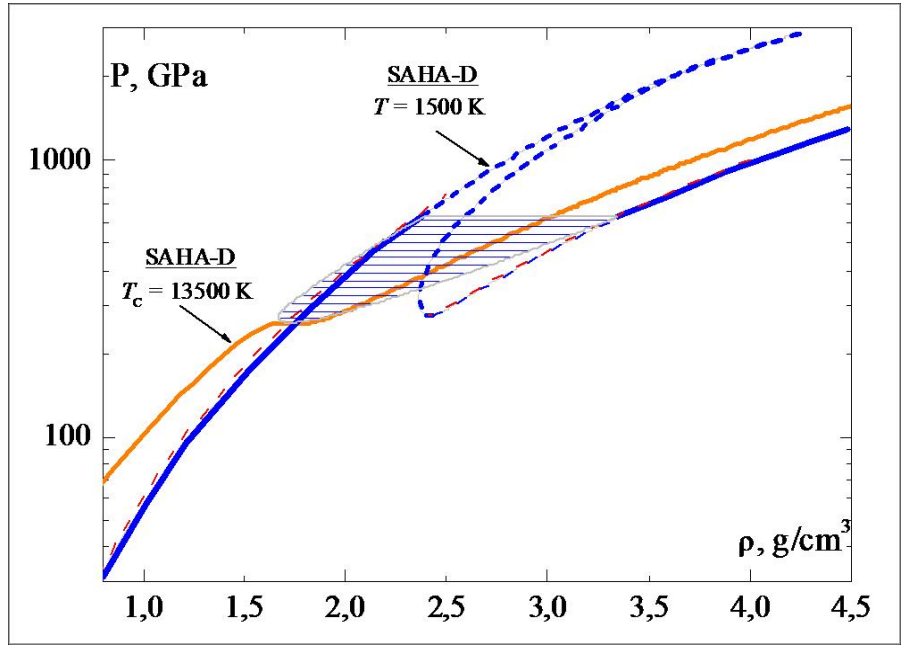

**Figure 8**. *P–ρ* phase diagram for hypothetical dissociation-driven phase transition in dense deuterium (SAHA-D code [21]). Solid lines – calculated isotherms for $T \ll T_c$, and $T = T_c \approx 13500$ K. Dashed curves at $T = 1500$ K – initial meta- and unstable parts of isotherm. Shaded area – two-phase region. (Figure from [21]).

## **Anomalous thermodynamics in the vicinity of two-phase region of entropic PTs**

It should be stressed also that negative sign of $(\partial P/\partial T)_V$ is always accompanied by negative sign of sequence of thermodynamic derivatives, which are positive usually. The most important is negativity of three derivatives: (*i*) – thermodynamic Gruneizen coefficient, i.e. $V(\partial P/\partial U)_V < 0$; (*ii*) – entropic pressure coefficient, i.e. $(\partial P/\partial S)_V < 0$, and (*iii*) – thermal expansion coefficient, i.e. $(\partial V/\partial T)_P < 0$ (here *U* and *S* – internal energy and entropy). Negativity of all notified above derivatives leads to very important consequences in mutual order and behavior of all thermodynamic isolines, i.e. isotherms, isentropes and shock adiabats first of all. It is of primary importance also for *hydrodynamics* of adiabatic flows, e.g. shock compression, isentropic expansion, adiabatic expansion into vacuum, spinodal decomposition etc. All these problems should be discussed separately [22] [23] (see also [24]).

Even more clearly anomalous thermodynamics in the vicinity of two-phase region for entropic PTs could be illustrated on another example – dissociation-driven phase transition in simplified EOS model for shock-compressed deuterium (Figure 8) (see also Figure 4 in [21]). Two anomalous features, exposed at this figure, should be emphasized in addition to those mentioned above. Namely, (*i*) – spinodal cupola, which is always located *inside* binodal cupola in the case of enthalpic VdW-like PT, now located partially *outside* of binodal area for entropic PT; (*ii*) – spinodal point of rare phase may have *higher density* than spinodal point of dense phase at low enough temperature. All mentioned above anomalies have clear geometric interpretation: - temperature, energy and entropy surfaces as functions of pressure and density, have *multi-layered* structure over the *P–V* plane in the case of all entropic phase transitions, in particular for ionization- and dissociation-driven PTs.

## What should we classify in case of unexplored phase transition [4][26]

Keeping in mind discussed above difference between enthalpic and entropic phase transitions we ought to summarize main features, which should be classified, when one meets unexplored phase transition:

- Is this PT of $1^{st}$ or $2^{nd}$-order?
- Is this PT enthalpic or entropic?
- Is this PT isostructural or non-isostructural?
- Is this PT congruent or non-congruent?
- Do we use Coulombless approximation in description of this PT, or we take into account all consequences of long-range nature of Coulomb interaction?
- What is the scenario of phase transformation in two-phase region? Is it macro- or mesoscopic one (structured mixed phase scenario ("pasta"))?

## Conclusions

- Widely accepted visible equivalence of gas-liquid-like and quark-hadron (deconfinement) phase transitions in high energy density nuclear matter is illusive.
- Both phase transitions belong to fundamentally different classes: Gas-Liquid PT is enthalpic one, while Quark-Hadron PT is entropic one.
- Properties of entropic and enthalpic PTs differ significantly from each other.
- Pressure-temperature dependence of phase boundary for enthalpic phase transition (HPT) and entropic one (SPT) have different sign $[(dP/dT)_{HPT} \geq 0 \,//\, (dP/dT)_{SPT} \leq 0]$.
- Isotherms of entropic PT have anomalous behavior within the two-phase region at sufficiently low temperature and anomalous order of metastable and unstable parts: e.g. stable-I / metastable-I / unstable-I / *metastable*-III / unstable-II / metastable-II / stable-II.
- Binodals and spinodals of entropic PT have anomalous order. Isothermal spinodal $[(\partial P/\partial V)_T = 0]$ may be located *outside* binodal of entropic PT at low enough temperature.
- Two-phase region and its close vicinity for entropic PT obey to anomalous thermodynamics. Namely: negative Gruneizen parameter, negative thermal and entropic pressure coefficients, negative thermal expansion coefficient *etc*. Besides there are anomalous order of isotherms, anomalous order of isentropes and anomalous order of shock adiabats *etc*.
- Deconfinement-driven (QHPT) and ionization-driven "plasma" phase transitions (PPT) as well as dissociation- and depolimerization-driven PTs in $N_2$ *etc*. are entropic PT, and hence they have many common features in spite of many order difference in density and energy.
- It is promising to investigate entropic PTs experimentally, for example via heavy ion beams (HIB) and surface laser heating *etc*. It is promising also to investigate entropic PTs theoretically in frames of traditional thermodynamic models (chemical picture) and via *ab initio* approaches especially.

**Acknowledgements**.

Author acknowledges V. Gryaznov for collaboration and obtaining phase diagram of deuterium, and M. Hempell and V. Dexheimer for collaboration and permission to use joint density-temperature phase diagram for GLPT and QHPT. Author acknowledges also V. Fortov, D. Blaschke and J. Randrup for helpful and fruitful discussions. This work was supported by the Presidium RAS Scientific Program "Physics of extreme states of matter" and partially by Extreme Matter Institute – EMMI (Germany).